\documentclass[twocolumn,showpacs,prl,superscriptaddress]{revtex4}

\usepackage{graphicx}
\usepackage{amsmath}
\usepackage{amssymb}

\begin{document}

\title{The role of magnetic polarons in ferromagnetic GdN}

\author{F.~Natali}
  \affiliation{The MacDiarmid Institute for Advanced Materials and Nanotechnology, School of Chemical and Physical Sciences, Victoria University of Wellington, PO Box 600, Wellington 6140, New Zealand}
\author{B.~J.~Ruck} \email{ben.ruck@vuw.ac.nz}
  \affiliation{The MacDiarmid Institute for Advanced Materials and Nanotechnology, School of Chemical and Physical Sciences, Victoria University of Wellington, PO Box 600, Wellington 6140, New Zealand}
\author{H.~J.~Trodahl}
  \affiliation{The MacDiarmid Institute for Advanced Materials and Nanotechnology, School of Chemical and Physical Sciences, Victoria University of Wellington, PO Box 600, Wellington 6140, New Zealand}
\author{Do~Le~Binh}
  \affiliation{The MacDiarmid Institute for Advanced Materials and Nanotechnology, School of Chemical and Physical Sciences, Victoria University of Wellington, PO Box 600, Wellington 6140, New Zealand}
\author{S.~Vezian}
  \affiliation{Centre de Recherche sur l'Hetero-Epitaxie et ses Applications (CRHEA), Centre National de la Recherche Scientifique, Rue Bernard Gregory, F-06560 Valbonne, France}
\author{B.~Damilano}
  \affiliation{Centre de Recherche sur l'Hetero-Epitaxie et ses Applications (CRHEA), Centre National de la Recherche Scientifique, Rue Bernard Gregory, F-06560 Valbonne, France}
\author{Y.~Cordier}
  \affiliation{Centre de Recherche sur l'Hetero-Epitaxie et ses Applications (CRHEA), Centre National de la Recherche Scientifique, Rue Bernard Gregory, F-06560 Valbonne, France}
\author{F.~Semond}
  \affiliation{Centre de Recherche sur l'Hetero-Epitaxie et ses Applications (CRHEA), Centre National de la Recherche Scientifique, Rue Bernard Gregory, F-06560 Valbonne, France}
\author{C.~Meyer}
  \affiliation{Institut Neel, Centre National de la Recherche Scientifique and Universite Joseph Fourier, B.P. 166, F-38042 Grenoble Cedex, France}

\begin{abstract}
We report an interplay between magnetism and charge transport in the ferromagnetic semiconductor GdN, pointing to the formation of magnetic polarons centred on nitrogen vacancies. The scenario goes some way to resolving a long-standing disagreement between the measured and predicted Curie temperature in GdN. It further constitutes an extension of concepts that relate closely to the behaviour of ferromagnetic semiconductors generally, and EuO in particular.
\end{abstract}

\pacs{75.50.Pp, 75.30.-m, 75.10.-b}
\date{\today}
\maketitle

Intrinsic ferromagnetic semiconductors, which offer the freedom to dope into specific conducting channels without destroying their magnetic behaviour, are of obvious technological interest. From a fundamental point of view they are again interesting; it is a substantial challenge to understand the interplay between their magnetic and transport properties. Among the earliest discovered, and even now the most studied, is EuO, which shows a magnetoresistance as large as thirteen orders of magnitude across the metal-insulator transition at the Curie temperature ($T_C$)~\cite{Shapira_Reed,Oliver_Reed,Penney_Torrance}, the largest in any compound. The strong transport-magnetism interplay is further emphasised by the enhanced $T_C$ seen in electron-doped samples~\cite{Mauger_Godart,Sutarto_Tjeng}. Debate continues about whether the ferromagnetic transition is homogeneous~\cite{Oliver_Reed,Sinjukow_Nolting,Mairoser_Mannhart} or whether it involves magnetic polarons nucleating around magnetic impurities~\cite{Torrance_McGuire,Penney_Torrance,Arnold_Kroha,Hillery_Liu,Snow_Ansermet,Liu_Tang}.

GdN, the prototypical rare-earth nitride compound, provides a rich set of comparisons with EuO. Divalent Eu and trivalent Gd share the same half-filled $4f$ shell, with $S=7/2$, $L=0$, and a net moment of $7~\mu_\mathrm{B}$. They share also the same rock salt structure, their reported $T_C$ are both near 70~K~\cite{Leuenberger_Hessler,Granville_Trodahl,Senapati_Barber,Scarpulla_Gossard}, and GdN also shows a strong magnetoresistance at $T_C$~\cite{Leuenberger_Hessler,Ludbrook_Durbin}. Theoretical treatments of GdN~\cite{Larson_Schilfgaarde,Duan_Tsymbal,Chantis_Kotani,Mitra_Lambrecht2} reproduce the measured electronic band features well, showing agreement as regards its semiconductor nature~\cite{Leuenberger_Hessler,Granville_Trodahl}, the direct band gap~\cite{Trodahl_Lambrecht,Yoshitomi_Sakurai,Yoshitomi_Sakurai2}, and features in the conduction-band density of states~\cite{Leuenberger_Hessler,Preston_Trodahl}. Even more important for device development is that GdN, unlike EuO, has a dispersive valence band of delocalised nitrogen states, which raises the possibility that it can support both $n-$ and $p-$type conduction. Its potential in realistic spintronics has already been demonstrated by its use in a spin filter~\cite{Senapati_Barber2}.

In contrast the ferromagnetic exchange mechanism is still poorly understood. The atomic-like nature of the $4f$ electrons necessitates indirect exchange, but those levels lie too far below the Fermi level to call on the third-order perturbation theory applied to EuO~\cite{Mauger_Godart,Kasuya,Lee_Liu}.  The validity of a proposed exchange channel involving the excited Gd $4f^8$ level is uncertain~\cite{Kasuya_Li}; in particular note these levels are similarly far from Fermi level~\cite{Preston_Trodahl}. Small energy differences are found among various spin orderings determined within the LDSA+$U$ treatment that has successfully reproduced the band structure, but they lead to an estimated $T_C$ below 25~K~\cite{Mitra_Lambrecht}, far below the experimental values. There is even a 1980 report~\cite{Wachter_Kaldis}, raised again recently~\cite{Wachter}, that GdN is metamagnetic and metallic rather than strictly ferromagnetic, though no recent data support that phase.

In view of the $T_C$ discrepancy, there have been several attempts to explore free-carrier exchange mechanisms in GdN. Metallic versions of the problem are in disagreement with transport measurements~\cite{Duan_Tsymbal,Duan_Tsymbal2}, but there is the potential that even in a semiconductor, Ruderman-Kittel-Kasuya-Yosida (RKKY) exchange might lead to an enhancement peaking at 60~K for electron doping near $10^{20}$~cm$^{-3}$~\cite{Sharma_Nolting}. Such a carrier concentration is easily obtained in the rare-earth nitrides, which show a propensity for nitrogen vacancy formation~\cite{Granville_Trodahl,Ludbrook_Durbin,Senapati_Barber}. However, $T_C$ does not peak; it is found to remain stable near 70~K for a very wide range of carrier concentrations~\cite{Plank_Ruck}, rising above that value only for very heavily doped material~\cite{Plank_Ruck,Senapati_Barber,Senapati_Barber3}.

\begin{figure}[h!]
  \centering
  \includegraphics[width=7.5cm]{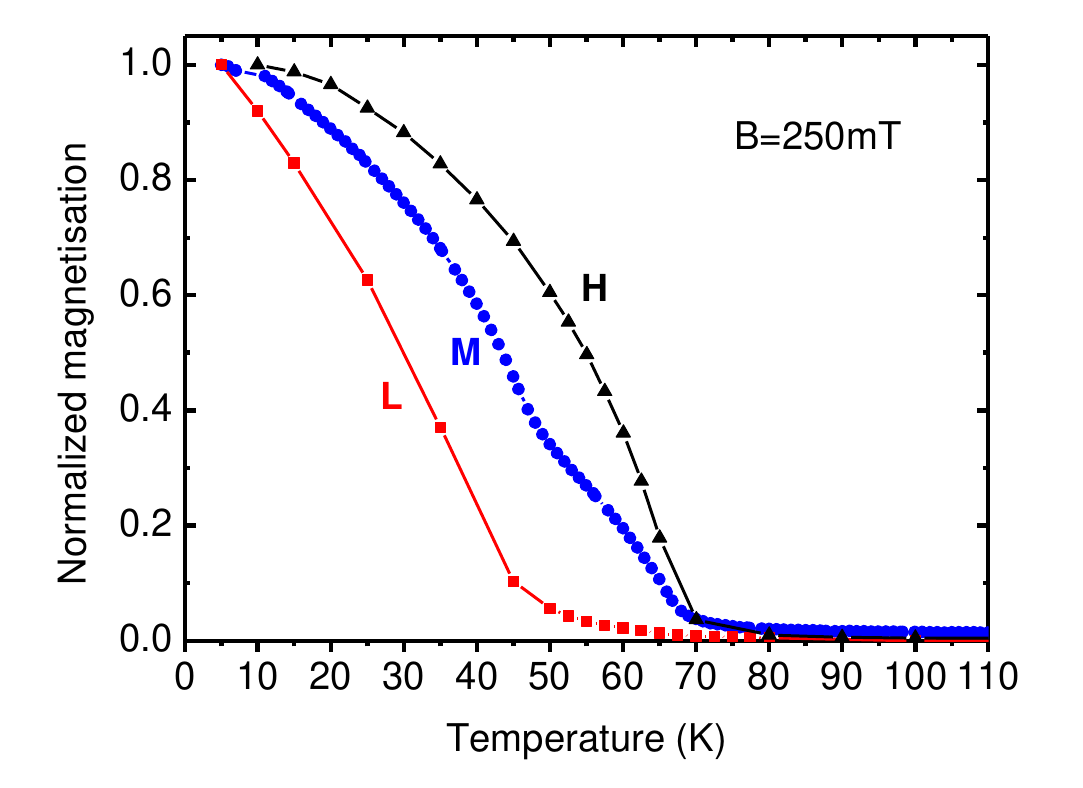}
  \caption{(color online) Temperature dependent magnetisation of GdN films with carrier concentrations of $<10^{18}$~cm$^{-3}$ (sample $L$), $3.3\times10^{20}$~cm$^{-3}$ (sample $M$), and $2\times10^{21}$~cm$^{-3}$ (sample $H$).}
  \label{MvsTfoot}
\end{figure}

In this letter we address the nature of the ferromagnetic transition in GdN using experimental data from a series of epitaxial films with systematically varied doping levels. The study is motivated by the existence in the literature of data showing a foot at the base of the magnetisation curve~\cite{Granville_Trodahl,Plank_Ruck,Yoshitomi_Sakurai}, data similar to those taken on recent films grown in our laboratory (Figure~\ref{MvsTfoot})~\cite{MvsTSamples}. In samples with heavy doping the magnetisation nucleates uniformly near 70~K (sample $H$ of Fig.~\ref{MvsTfoot}), which we will show is aided by mobile charge carriers, but it drops to another homogeneous transition near 50~K in samples with very low doping (sample $L$). Samples of intermediate doping (sample $M$) still show a transition to a ferromagnetic state near the 70~K $T_C$ reported so often in the literature~\cite{Leuenberger_Hessler,Granville_Trodahl,Natali_Hirsch,Ludbrook_Durbin,Senapati_Barber,Scarpulla_Gossard}, but the transition is incomplete until the temperature falls below $\sim$50~K. Reduced doping then results in a smaller magnetisation between 70 and 50~K, rather than the reduced $T_C$ as expected for a homogeneous carrier-mediated drift in the exchange interaction strength.

\begin{table}
\begin{tabular}{|c|c|c|c|}
\hline
  \hline
Sample &  Rocking curve & $n$ (300~K)\\
 &  FWHM ($^\circ$) & (cm$^{-3}$)\\
  \hline
  $M$ &  2.36  & $3.3 \times10^{20}$\\
  $M_1$ &  2.37  & $4.4 \times10^{20}$\\
  $M_2$ &  2.64  & $5.7 \times10^{20}$\\
  \hline
  \hline
\end{tabular}
\caption{Structure and transport properties of the three GdN films shown in Fig.~\ref{MvsTvsH}. The rocking curve was performed on the symmetric (111) GdN x-ray diffraction peak.} \label{Props}
\end{table}
It is significant that films grown at temperatures permitting epitaxial growth lie in the high-conductivity class (sample $H$ of Fig.~\ref{MvsTfoot}); lower-conductivity films have been reported only for ambient-temperature grown polycrystalline films. Thus to explore the situation further we focus here on three well-ordered epitaxial films (including sample $M$) that we have managed to grow, under careful temperature control, with carrier concentrations between samples $L$ and $H$ in Fig.~\ref{MvsTfoot}. We have investigated the magnetic and conducting state of the films, including Hall effect measurements to obtain the carrier concentration ($n$) and the mobility ($\mu=1/(en\rho)$ with $\rho$ the resistivity and $e$ the electron's charge) as a function of temperature. The full data set shows it is magnetic polarons centred on charged nitrogen vacancies that nucleate ferromagnetism at 70~K.

The three samples ($M$, $M_1$, $M_2$) were grown in a Riber molecular beam epitaxy system using NH$_3$ as the nitrogen precursor and a Gd solid effusion cell. The 120~nm thick GdN films were deposited onto 100~nm thick (0001) AlN layers which were themselves grown on (111) silicon substrates. Tuning the growth temperature of the GdN films between 650$^\circ$C and 670$^\circ$C allowed preparation of samples with slightly varied residual doping (see Table~\ref{Props}). The GdN growth was always under N-rich conditions, with a beam equivalent pressure of $1.9\times10^{-5}$~Torr and $5\times10^{-8}$~Torr for NH$_3$ and Gd, respectively, which leads to a growth rate of $0.12\pm0.01~\mu$m/h. The epitaxial nature of the films was confirmed during growth by reflection high-energy electron diffraction. The GdN layers were capped with GaN to prevent decomposition in air. X-ray diffraction measurements show that the crystallinity is similar for each sample, with the GdN fully (111) oriented and displaying 6-fold symmetry as expected for its growth on a (0001) wurtzite surface. All films are fully relaxed and no evidence of impurity phases is found. Table~\ref{Props} also presents the carrier concentration of each film at 300~K, measured by the Hall effect. In each case the carriers are electrons, as expected for carriers originating from nitrogen vacancies.

\begin{figure}[h!]
  \centering
  \includegraphics[width=8cm]{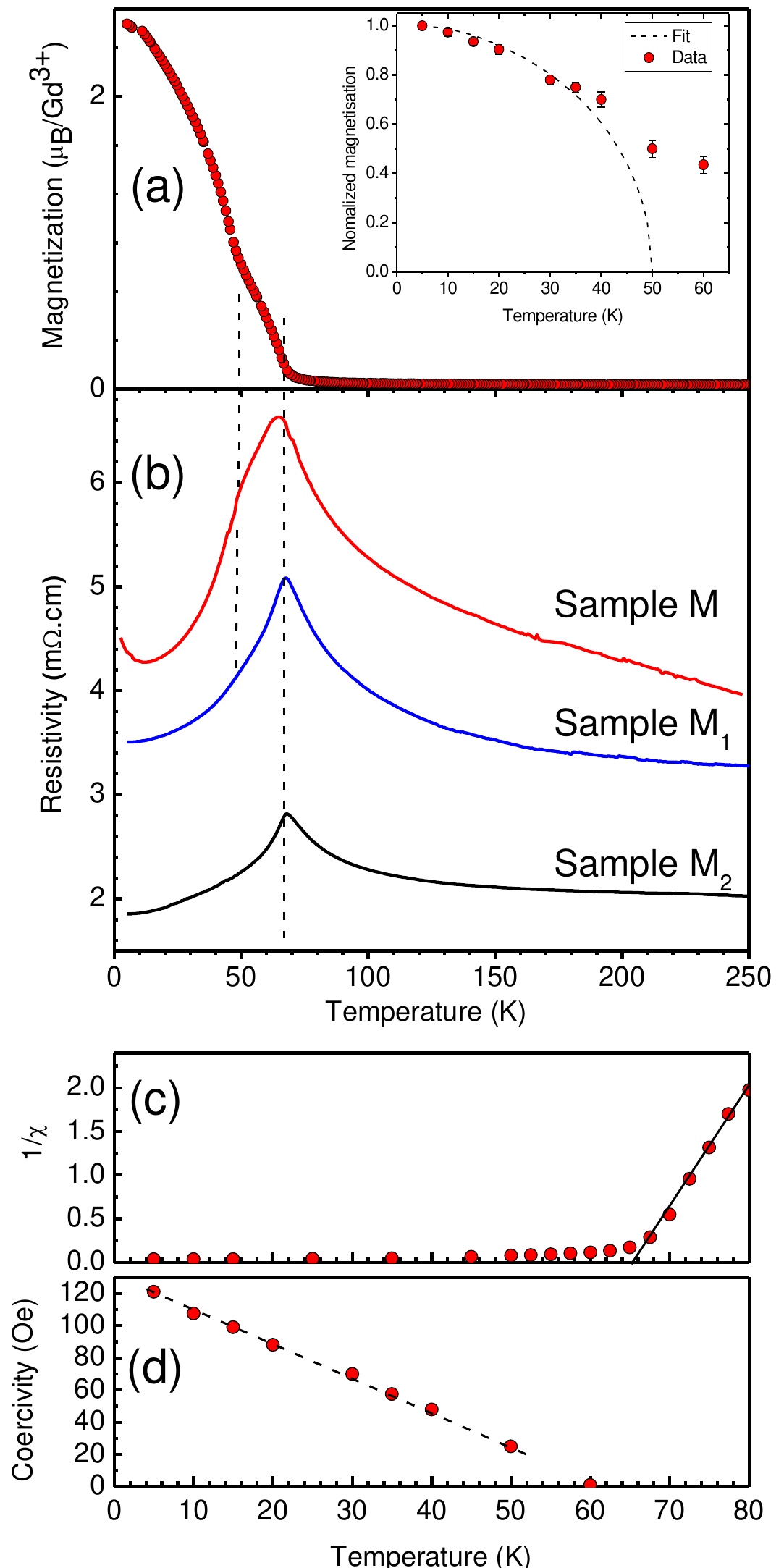} 
  \caption{(color online) (a)~Field-cooled temperature dependent magnetisation of sample M. Inset:~high-field remanant magnetisation. The dashed curve is a fit of the low temperature data to a ferromagnetic Brillouin function, yielding an intrinsic $T_C$ of only 50~K. (b)~Temperature dependent resistivity of GdN films with slightly varied carrier concentrations. (c)~Inverse susceptibility versus temperature indicating a paramagnetic Curie temperature of 65~K. (d)~Coercive field versus temperature.}
  \label{MvsTvsH}
\end{figure}

The in-plane field-cooled magnetisation of sample $M$ under an applied field of 250~Oe is shown in Figure~\ref{MvsTvsH}(a). The magnetisation shows an upturn below $\approx\,$65~K, matching the Curie temperature $T_C$ of 65$\pm$1~K deduced from the inverse of susceptibility in the paramagnetic regime [Fig.~\ref{MvsTvsH}(c)]. In addition, there is a clear shoulder in the magnetisation at about 50~K, coinciding with the separation of the field cooled and zero field cooled curves (not shown). Field dependent magnetisation measurements confirm the vanishing of hysteresis above 50~K [Fig.~\ref{MvsTvsH}(d)]. Films $M_1$ and $M_2$ both show similar behaviour, although with less pronounced shoulders at 50~K. As emphasised in Fig.~\ref{MvsTfoot} the shoulder is not observed in more conducting samples~\cite{Ludbrook_Durbin,Natali_Hirsch}. It is worth mentioning that this shoulder is strongly reminiscent of the ``double-dome'' seen in the magnetisation of electron-doped EuO~\cite{Liu_Tang,Mauger_Godart}.

The temperature-dependent saturation magnetisation is plotted in the inset to Fig.~\ref{MvsTvsH}(a), with the data at low temperature fitted by the expected Brillouin function for ($S=7/2$)~\cite{Blundell}. Remarkably this extrapolates to a Curie temperature of 50~K, substantially less than the 65~K temperature at which magnetic ordering sets in, but coincident with the knee in the magnetisation. The clear implication is that the ferromagnetic ordering that sets in near 65~K differs from the homogeneous low temperature magnetic phase.

Turning to electron transport, the corresponding temperature-dependent resistivities are shown in Fig.~\ref{MvsTvsH}(b). Both the room-temperature values and the temperature dependencies are in agreement with previous reports~\cite{Leuenberger_Hessler,Granville_Trodahl}, with a trend that is consistent with the carrier concentrations in Table ~\ref{Props}. The data lie between the extremes of resistivity in the literature, in which less resistive samples~\cite{Ludbrook_Durbin,Scarpulla_Gossard} show a positive temperature coefficient of resistance (TCR) near room temperature, while more resistive samples typically show a strong negative TCR~\cite{Granville_Trodahl,Plank_Ruck}. All show a peak near 70~K, but note that for the most resistive samples in the literature ($\sim10~\Omega$cm at 300~K) the peak shifts downward to about 50~K~\cite{Plank_Ruck}.

The present three films appear to lie right at the apparent metal-insulator transition found in the literature. Film $M_2$, with the highest electron concentration of the three samples, has a very small negative TCR at room temperature and has a metallic positive TCR from $T_C$ down to 5~K. At lower electron concentrations ($M$, $M_1$) the negative TCR at 300~K is sequentially stronger and the amplitude of the peak near $T_C$ increases. The most resistive sample $M$ shows a negative TCR at the lowest temperatures indicating that the ferromagnetic phase is also semiconducting. Furthermore, a shoulder appears in the resistivity below $T_C$ at about 50~K that is not seen clearly in the other two films. The appearance of shoulders near 50~K in both the resistivity and the magnetisation clearly links the charge carriers and magnetic exchange mechanism. This systematic behaviour is reminiscent of that observed in Eu-rich EuO~\cite{Oliver_Reed}, although we note that the resistivity peak is substantially sharper in the present GdN samples than in EuO of comparable resistivity.

It is notable that the resistivity of film $M$ bears a similarity to the temperature dependence discussed many years ago for magnetic metals, with an elbow-like anomaly on the low-temperature side of a resistivity peak. The anomaly generates a peak in the derivative $d\rho/dT$ that is identified as occurring at $T_C$~\cite{Fisher_Langer}. The more rounded peak in the resistance at higher temperatures signals a point at which the spin-spin correlation length $\sigma$ satisfies $1/\sigma\approx2k_F$, where $k_F$ is the Fermi wave radius. It is however clear that the present system, GdN, cannot be interpreted on that basis on two grounds: (i)~it is semiconducting without a well-defined Fermi radius and (ii)~the magnetisation indicates a paramagnetic Curie temperature at the resistivity peak, rather than the derivative peak. We thus seek an explanation within an inhomogeneous model.

In view of the dominant doping role played by nitrogen vacancies ($V_\mathrm{N}$) it is here that we seek an enhanced exchange interaction that may nucleate ferromagnetism at 70~K. As a guide we use the $V_\mathrm{N}$ energy levels calculated by Punya \textit{et al.}~\cite{Punya_Lambrecht}; each vacancy is expected to bind two electrons in a singlet state, with a third electron bound only weakly below the conduction band in the dilute limit. As the temperature lowers the carriers begin to freeze out of the conduction band to occupy the third $V_\mathrm{N}$ level. This lends a magnetic moment of $1~\mu_B$ to the vacancies but more importantly provides a large local electron density that can mediate exchange between neighbouring Gd ions within the localised cloud, thus forming a magnetic polaron~\cite{Coey_Venkatesan}. Evidently ferromagnetism first nucleates around these sites at a temperature of 65-70~K, forming a nonuniform magnetic phase with the vanishing coercivity seen in Fig.~\ref{MvsTvsH}(d).

The most lightly doped of the present samples ($M$) shows an activation that would place the third $V_\mathrm{N}$ level about 6~meV below the conduction band (see below), though more resistive films in the literature would place it somewhat deeper~\cite{Plank_Ruck}. The difference is likely due to the high $V_\mathrm{N}$ density in these films; note that the carrier density of film $M$ corresponds to one electron per 100 unit cells, separated by $\sim1.5$~nm, very close to the $\sim1.4$~nm radius of the Bohr orbit obtained using an effective mass $m^*=0.15m_0$ estimated from the band structure~\cite{Trodahl_Lambrecht} and  $\epsilon = 4$ measured by the subgap
reflectivity~\cite{AzeemPComm}. It is significant that this is just the $V_\mathrm{N}$ density at which both the metal-insulator and the 70-50~K ferromagnetic transitions occur. Thus at higher carrier concentrations, such as in film $M_2$, the carriers overlap and the ferromagnetic phase forms uniformly near 70~K. Even higher doping levels can mediate a $T_C$ enhancement above 70~K~\cite{Plank_Ruck,Senapati_Barber,Senapati_Barber3}.

\begin{figure}[h!]
  \centering
  \includegraphics[width=7.5cm]{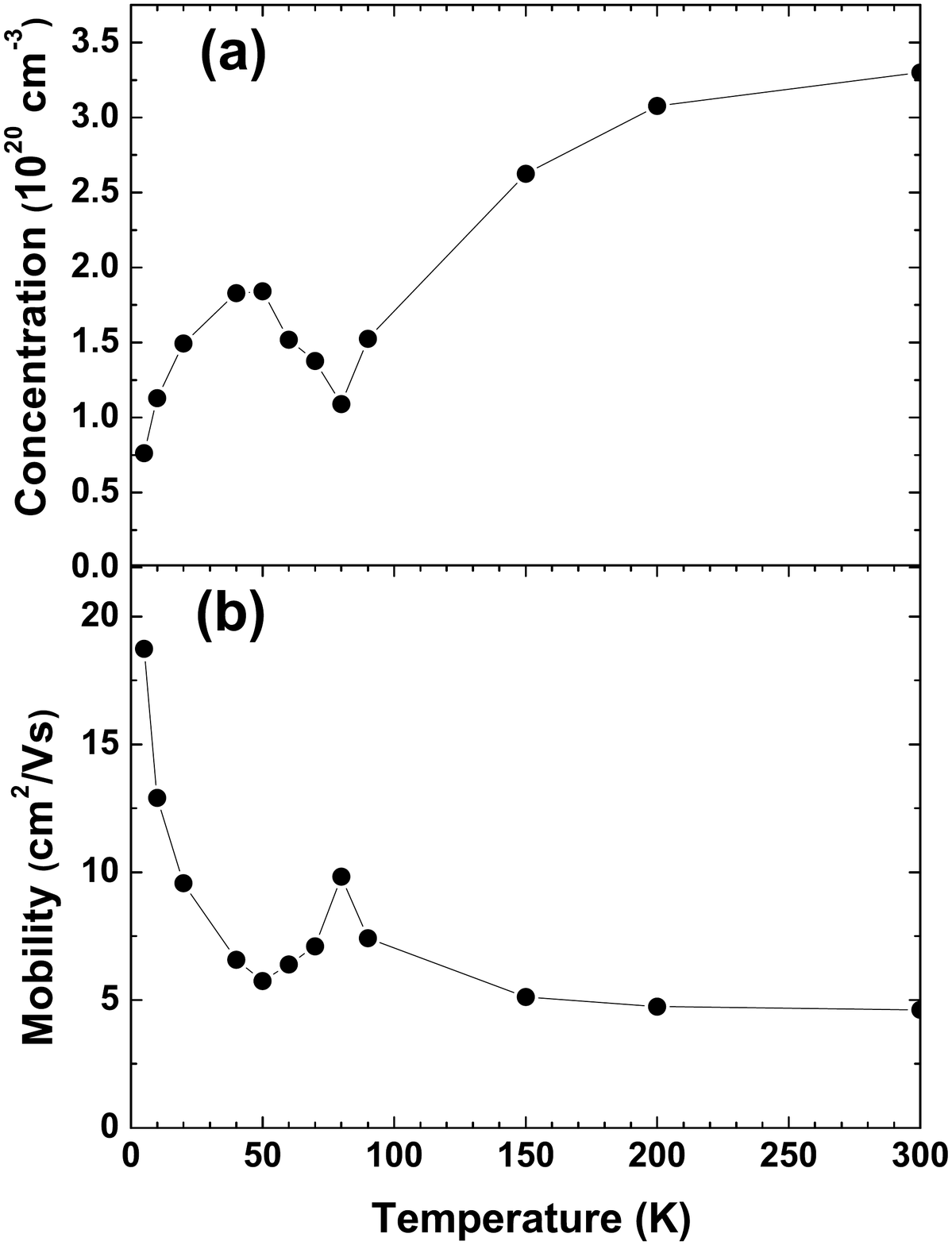}
  \caption{(color online) Temperature dependence of (a)~the electron concentration $n$ and (b)~the mobility $\mu$ for sample $M$, obtained from the high-field Hall effect.}
  \label{HallRes}
\end{figure}

The discussion above emphasises the importance of the charge carrier dynamics in determining both the transport and magnetic properties of GdN. We have thus followed the carrier concentration as a function of the temperature by measuring the Hall resistance $R_H$. The carrier concentration, $n$, is then signalled by the slope at fields sufficiently strong to saturate the magnetization, and thus also the extraordinary Hall effect. The ordinary Hall effect then leads to a high-field slope that measures a carrier density that is very nearly independent of temperature in the most conductive film $M_2$, placing this film on the metallic side of the metal-insulator transition. In contrast the most weakly doped film $M$ shows a high-field slope that implies the semiconductor-like temperature-dependent carrier density plotted in Figure~\ref{HallRes}(a). The carriers are activated above 65~K, with an activation energy near 6~meV, confirming the semiconductor nature of the paramagnetic phase. The freeze-out is halted at 70~K, as the gap closes and the activation energy falls in the ferromagnetic near-vacancy regions, but resumes with a reduced activation energy of 1~meV below the 50~K transition to a homogeneous ferromagnetic phase.

The temperature-dependent mobility plotted in Figure~\ref{HallRes}(b) is quite low in comparison with conventional semiconductors, and remarkable for (i)~the temperature independance above 100~K, and (ii)~the rapid rise below the Curie temperature. The implication is that electron scattering is dominated by the magnetic disorder; in particular static impurities would continue to limit the mobility at low temperature in disagreement with the data. he mobility decreases on entry into the inhomogeneous magnetic phase below $\sim70$~K, and there is a mobility minimum (scattering cross section maximum) at 50~K, exactly as is expected at the Curie temperature for spin-disorder scattering~\cite{deGennes_Friedel,Fisher_Langer}. The resistive anomaly near 65~K thus has contributions from both spin disorder scattering and a closing majority-spin gap in the ferromagnetic state~\cite{Trodahl_Lambrecht}.

Now summarising the main results of this experimental investigation, we have examined the magnetic response and charge transport in GdN with carrier concentrations across the metal-insulator transition. The data show clear evidence of a double ferromagnetic transition, with a homogeneous ferromagnetic phase only below 50~K. The remarkable interplay between magnetism and charge carriers in GdN, and especially its doping dependence, signals the presence of magnetic polarons that nucleate about nitrogen vacancies. The polarons render a major volume fraction ferromagnetic at even such a small vacancy concentration as 1\%, and lead to the commonly quoted 65-70~K Curie temperature. Stoichiometric GdN then has a Curie temperature near 50~K, in closer agreement with theoretical predictions. This scenario is essential for understanding the distinct ``double-dome'' shape of the temperature-dependent resistivity and magnetisation.  Interestingly, our results have similarity in many crucial aspects with the most studied ferromagnetic semiconductor EuO, where closely related concepts, and particularly the effects of magnetic polarons, continue to be discussed.

\begin{acknowledgments}
We acknowledge funding from the NZ FRST (VICX0808) and the Marsden Fund (08-VUW-030). We thank S.~Granville and S.V.~Chong from IRL for technical assistance during magnetic measurements, and T.~Maity for providing the data for sample $L$.
\end{acknowledgments}

\bibliography{references}

\begin{thebibliography}{44}
\expandafter\ifx\csname natexlab\endcsname\relax\def\natexlab#1{#1}\fi
\expandafter\ifx\csname bibnamefont\endcsname\relax
  \def\bibnamefont#1{#1}\fi
\expandafter\ifx\csname bibfnamefont\endcsname\relax
  \def\bibfnamefont#1{#1}\fi
\expandafter\ifx\csname citenamefont\endcsname\relax
  \def\citenamefont#1{#1}\fi
\expandafter\ifx\csname url\endcsname\relax
  \def\url#1{\texttt{#1}}\fi
\expandafter\ifx\csname urlprefix\endcsname\relax\def\urlprefix{URL }\fi
\providecommand{\bibinfo}[2]{#2}
\providecommand{\eprint}[2][]{\url{#2}}

\bibitem[{\citenamefont{Shapira et~al.}(1973)\citenamefont{Shapira, Foner, and
  Reed}}]{Shapira_Reed}
\bibinfo{author}{\bibfnamefont{Y.}~\bibnamefont{Shapira}},
  \bibinfo{author}{\bibfnamefont{S.}~\bibnamefont{Foner}}, \bibnamefont{and}
  \bibinfo{author}{\bibfnamefont{T.~B.} \bibnamefont{Reed}},
  \bibinfo{journal}{Phys.\ Rev.\ B} \textbf{\bibinfo{volume}{8}},
  \bibinfo{pages}{2299} (\bibinfo{year}{1973}).

\bibitem[{\citenamefont{Oliver et~al.}(1972)\citenamefont{Oliver, Dimmock,
  McWhorter, and Reed}}]{Oliver_Reed}
\bibinfo{author}{\bibfnamefont{M.~R.} \bibnamefont{Oliver}},
  \bibinfo{author}{\bibfnamefont{J.~O.} \bibnamefont{Dimmock}},
  \bibinfo{author}{\bibfnamefont{A.~L.} \bibnamefont{McWhorter}},
  \bibnamefont{and} \bibinfo{author}{\bibfnamefont{T.~B.} \bibnamefont{Reed}},
  \bibinfo{journal}{Phys.\ Rev.\ B} \textbf{\bibinfo{volume}{5}},
  \bibinfo{pages}{1078} (\bibinfo{year}{1972}).

\bibitem[{\citenamefont{Penney et~al.}(1972)\citenamefont{Penney, Shafer, and
  Torrance}}]{Penney_Torrance}
\bibinfo{author}{\bibfnamefont{T.}~\bibnamefont{Penney}},
  \bibinfo{author}{\bibfnamefont{M.~W.} \bibnamefont{Shafer}},
  \bibnamefont{and} \bibinfo{author}{\bibfnamefont{J.~B.}
  \bibnamefont{Torrance}}, \bibinfo{journal}{Phys.\ Rev.\ B}
  \textbf{\bibinfo{volume}{5}}, \bibinfo{pages}{3669} (\bibinfo{year}{1972}).

\bibitem[{\citenamefont{Mauger and Godart}(1986)}]{Mauger_Godart}
\bibinfo{author}{\bibfnamefont{A.}~\bibnamefont{Mauger}} \bibnamefont{and}
  \bibinfo{author}{\bibfnamefont{C.}~\bibnamefont{Godart}},
  \bibinfo{journal}{Phys.\ Rep.} \textbf{\bibinfo{volume}{141}},
  \bibinfo{pages}{51} (\bibinfo{year}{1986}).

\bibitem[{\citenamefont{Sutarto et~al.}(2009)\citenamefont{Sutarto, Altendorf,
  Coloru, Moretti~Sala, Haupricht, Chang, Hu, Sch\"{u}{\ss}ler-Langeheine,
  Hollmann, Kierspel et~al.}}]{Sutarto_Tjeng}
\bibinfo{author}{\bibfnamefont{R.}~\bibnamefont{Sutarto}},
  \bibinfo{author}{\bibfnamefont{S.~G.} \bibnamefont{Altendorf}},
  \bibinfo{author}{\bibfnamefont{B.}~\bibnamefont{Coloru}},
  \bibinfo{author}{\bibfnamefont{M.}~\bibnamefont{Moretti~Sala}},
  \bibinfo{author}{\bibfnamefont{T.}~\bibnamefont{Haupricht}},
  \bibinfo{author}{\bibfnamefont{C.~F.} \bibnamefont{Chang}},
  \bibinfo{author}{\bibfnamefont{Z.}~\bibnamefont{Hu}},
  \bibinfo{author}{\bibfnamefont{C.}~\bibnamefont{Sch\"{u}{\ss}ler-Langeheine}},
  \bibinfo{author}{\bibfnamefont{N.}~\bibnamefont{Hollmann}},
  \bibinfo{author}{\bibfnamefont{H.}~\bibnamefont{Kierspel}},
  \bibnamefont{et~al.}, \bibinfo{journal}{Phys. Rev. B}
  \textbf{\bibinfo{volume}{80}}, \bibinfo{pages}{085308}
  (\bibinfo{year}{2009}).

\bibitem[{\citenamefont{Sinjukow and Nolting}(2003)}]{Sinjukow_Nolting}
\bibinfo{author}{\bibfnamefont{P.}~\bibnamefont{Sinjukow}} \bibnamefont{and}
  \bibinfo{author}{\bibfnamefont{W.}~\bibnamefont{Nolting}},
  \bibinfo{journal}{Phys.\ Rev.\ B} \textbf{\bibinfo{volume}{68}},
  \bibinfo{pages}{125107} (\bibinfo{year}{2003}).

\bibitem[{\citenamefont{Mairoser et~al.}(2010)\citenamefont{Mairoser, Schmehl,
  Melville, Heeg, Canella, B\"{o}ni, Zander, Schubert, Shai, Monkman
  et~al.}}]{Mairoser_Mannhart}
\bibinfo{author}{\bibfnamefont{T.}~\bibnamefont{Mairoser}},
  \bibinfo{author}{\bibfnamefont{A.}~\bibnamefont{Schmehl}},
  \bibinfo{author}{\bibfnamefont{A.}~\bibnamefont{Melville}},
  \bibinfo{author}{\bibfnamefont{T.}~\bibnamefont{Heeg}},
  \bibinfo{author}{\bibfnamefont{L.}~\bibnamefont{Canella}},
  \bibinfo{author}{\bibfnamefont{P.}~\bibnamefont{B\"{o}ni}},
  \bibinfo{author}{\bibfnamefont{W.}~\bibnamefont{Zander}},
  \bibinfo{author}{\bibfnamefont{J.}~\bibnamefont{Schubert}},
  \bibinfo{author}{\bibfnamefont{D.~E.} \bibnamefont{Shai}},
  \bibinfo{author}{\bibfnamefont{E.~J.} \bibnamefont{Monkman}},
  \bibnamefont{et~al.}, \bibinfo{journal}{Phys. Rev. Lett.}
  \textbf{\bibinfo{volume}{105}}, \bibinfo{pages}{257206}
  (\bibinfo{year}{2010}).

\bibitem[{\citenamefont{Torrance et~al.}(1972)\citenamefont{Torrance, Shafer,
  and McGuire}}]{Torrance_McGuire}
\bibinfo{author}{\bibfnamefont{J.~B.} \bibnamefont{Torrance}},
  \bibinfo{author}{\bibfnamefont{M.~W.} \bibnamefont{Shafer}},
  \bibnamefont{and} \bibinfo{author}{\bibfnamefont{T.~R.}
  \bibnamefont{McGuire}}, \bibinfo{journal}{Phys. Rev. Lett.}
  \textbf{\bibinfo{volume}{29}}, \bibinfo{pages}{1168} (\bibinfo{year}{1972}).

\bibitem[{\citenamefont{Arnold and Kroha}(2008)}]{Arnold_Kroha}
\bibinfo{author}{\bibfnamefont{M.}~\bibnamefont{Arnold}} \bibnamefont{and}
  \bibinfo{author}{\bibfnamefont{J.}~\bibnamefont{Kroha}},
  \bibinfo{journal}{Phys.\ Rev.\ Lett.} \textbf{\bibinfo{volume}{100}},
  \bibinfo{pages}{046404} (\bibinfo{year}{2008}).

\bibitem[{\citenamefont{Hillery et~al.}(1988)\citenamefont{Hillery, Emin, and
  Liu}}]{Hillery_Liu}
\bibinfo{author}{\bibfnamefont{M.~S.} \bibnamefont{Hillery}},
  \bibinfo{author}{\bibfnamefont{D.}~\bibnamefont{Emin}}, \bibnamefont{and}
  \bibinfo{author}{\bibfnamefont{N.-L.~H.} \bibnamefont{Liu}},
  \bibinfo{journal}{Phys. Rev. B} \textbf{\bibinfo{volume}{38}},
  \bibinfo{pages}{9771} (\bibinfo{year}{1988}).

\bibitem[{\citenamefont{Snow et~al.}(2001)\citenamefont{Snow, Cooper, Young,
  Fisk, Comment, and Ansermet}}]{Snow_Ansermet}
\bibinfo{author}{\bibfnamefont{C.~S.} \bibnamefont{Snow}},
  \bibinfo{author}{\bibfnamefont{S.~L.} \bibnamefont{Cooper}},
  \bibinfo{author}{\bibfnamefont{D.~P.} \bibnamefont{Young}},
  \bibinfo{author}{\bibfnamefont{Z.}~\bibnamefont{Fisk}},
  \bibinfo{author}{\bibfnamefont{A.}~\bibnamefont{Comment}}, \bibnamefont{and}
  \bibinfo{author}{\bibfnamefont{J.-P.} \bibnamefont{Ansermet}},
  \bibinfo{journal}{Phys. Rev. B} \textbf{\bibinfo{volume}{64}},
  \bibinfo{pages}{174412} (\bibinfo{year}{2001}).

\bibitem[{\citenamefont{Liu and Tang}(2012)}]{Liu_Tang}
\bibinfo{author}{\bibfnamefont{P.}~\bibnamefont{Liu}} \bibnamefont{and}
  \bibinfo{author}{\bibfnamefont{J.}~\bibnamefont{Tang}},
  \bibinfo{journal}{Phys. Rev. B} \textbf{\bibinfo{volume}{85}},
  \bibinfo{pages}{224417} (\bibinfo{year}{2012}).

\bibitem[{\citenamefont{Leuenberger et~al.}(2005)\citenamefont{Leuenberger,
  Parge, Felsch, Fauth, and Hessler}}]{Leuenberger_Hessler}
\bibinfo{author}{\bibfnamefont{F.}~\bibnamefont{Leuenberger}},
  \bibinfo{author}{\bibfnamefont{A.}~\bibnamefont{Parge}},
  \bibinfo{author}{\bibfnamefont{W.}~\bibnamefont{Felsch}},
  \bibinfo{author}{\bibfnamefont{K.}~\bibnamefont{Fauth}}, \bibnamefont{and}
  \bibinfo{author}{\bibfnamefont{M.}~\bibnamefont{Hessler}},
  \bibinfo{journal}{Phys.\ Rev.\ B} \textbf{\bibinfo{volume}{72}},
  \bibinfo{pages}{014427} (\bibinfo{year}{2005}).

\bibitem[{\citenamefont{Granville et~al.}(2006)\citenamefont{Granville, Ruck,
  Budde, Koo, Pringle, Kuchler, Bittar, Williams, and
  Trodahl}}]{Granville_Trodahl}
\bibinfo{author}{\bibfnamefont{S.}~\bibnamefont{Granville}},
  \bibinfo{author}{\bibfnamefont{B.~J.} \bibnamefont{Ruck}},
  \bibinfo{author}{\bibfnamefont{F.}~\bibnamefont{Budde}},
  \bibinfo{author}{\bibfnamefont{A.}~\bibnamefont{Koo}},
  \bibinfo{author}{\bibfnamefont{D.}~\bibnamefont{Pringle}},
  \bibinfo{author}{\bibfnamefont{F.}~\bibnamefont{Kuchler}},
  \bibinfo{author}{\bibfnamefont{A.}~\bibnamefont{Bittar}},
  \bibinfo{author}{\bibfnamefont{G.~V.~M.} \bibnamefont{Williams}},
  \bibnamefont{and} \bibinfo{author}{\bibfnamefont{H.~J.}
  \bibnamefont{Trodahl}}, \bibinfo{journal}{Phys. Rev. B}
  \textbf{\bibinfo{volume}{73}}, \bibinfo{pages}{235335}
  (\bibinfo{year}{2006}).

\bibitem[{\citenamefont{Senapati
  et~al.}(2011{\natexlab{a}})\citenamefont{Senapati, Fix, Vickers, Blamire, and
  Barber}}]{Senapati_Barber}
\bibinfo{author}{\bibfnamefont{K.}~\bibnamefont{Senapati}},
  \bibinfo{author}{\bibfnamefont{T.}~\bibnamefont{Fix}},
  \bibinfo{author}{\bibfnamefont{M.~E.} \bibnamefont{Vickers}},
  \bibinfo{author}{\bibfnamefont{M.~G.} \bibnamefont{Blamire}},
  \bibnamefont{and} \bibinfo{author}{\bibfnamefont{Z.~H.}
  \bibnamefont{Barber}}, \bibinfo{journal}{Phys. Rev. B}
  \textbf{\bibinfo{volume}{83}}, \bibinfo{pages}{014403}
  (\bibinfo{year}{2011}{\natexlab{a}}).

\bibitem[{\citenamefont{Scarpulla et~al.}(2009)\citenamefont{Scarpulla,
  Gallinat, Mack, Speck, and Gossard}}]{Scarpulla_Gossard}
\bibinfo{author}{\bibfnamefont{M.~A.} \bibnamefont{Scarpulla}},
  \bibinfo{author}{\bibfnamefont{C.~S.} \bibnamefont{Gallinat}},
  \bibinfo{author}{\bibfnamefont{W.~S.} \bibnamefont{Mack}},
  \bibinfo{author}{\bibfnamefont{J.~S.} \bibnamefont{Speck}}, \bibnamefont{and}
  \bibinfo{author}{\bibfnamefont{A.~C.} \bibnamefont{Gossard}},
  \bibinfo{journal}{J.\ Cryst.\ Growth} \textbf{\bibinfo{volume}{311}},
  \bibinfo{pages}{1239} (\bibinfo{year}{2009}).

\bibitem[{\citenamefont{Ludbrook et~al.}(2009)\citenamefont{Ludbrook, Farrell,
  Kuebel, Ruck, Preston, Trodahl, Ranno, Reeves, and Durbin}}]{Ludbrook_Durbin}
\bibinfo{author}{\bibfnamefont{B.~M.} \bibnamefont{Ludbrook}},
  \bibinfo{author}{\bibfnamefont{I.~L.} \bibnamefont{Farrell}},
  \bibinfo{author}{\bibfnamefont{M.}~\bibnamefont{Kuebel}},
  \bibinfo{author}{\bibfnamefont{B.~J.} \bibnamefont{Ruck}},
  \bibinfo{author}{\bibfnamefont{A.~R.~H.} \bibnamefont{Preston}},
  \bibinfo{author}{\bibfnamefont{H.~J.} \bibnamefont{Trodahl}},
  \bibinfo{author}{\bibfnamefont{L.}~\bibnamefont{Ranno}},
  \bibinfo{author}{\bibfnamefont{R.~J.} \bibnamefont{Reeves}},
  \bibnamefont{and} \bibinfo{author}{\bibfnamefont{S.~M.}
  \bibnamefont{Durbin}}, \bibinfo{journal}{J. Appl. Phys.}
  \textbf{\bibinfo{volume}{106}}, \bibinfo{pages}{063910}
  (\bibinfo{year}{2009}).

\bibitem[{\citenamefont{Larson et~al.}(2007)\citenamefont{Larson, Lambrecht,
  Chantis, and van Schilfgaarde}}]{Larson_Schilfgaarde}
\bibinfo{author}{\bibfnamefont{P.}~\bibnamefont{Larson}},
  \bibinfo{author}{\bibfnamefont{W.~R.~L.} \bibnamefont{Lambrecht}},
  \bibinfo{author}{\bibfnamefont{A.}~\bibnamefont{Chantis}}, \bibnamefont{and}
  \bibinfo{author}{\bibfnamefont{M.}~\bibnamefont{van Schilfgaarde}},
  \bibinfo{journal}{Phys.\ Rev.\ B} \textbf{\bibinfo{volume}{75}},
  \bibinfo{pages}{045114} (\bibinfo{year}{2007}).

\bibitem[{\citenamefont{Duan et~al.}(2007)\citenamefont{Duan, Sabiryanov, Mei,
  Dowben, Jaswal, and Tsymbal}}]{Duan_Tsymbal}
\bibinfo{author}{\bibfnamefont{C.-G.} \bibnamefont{Duan}},
  \bibinfo{author}{\bibfnamefont{R.~F.} \bibnamefont{Sabiryanov}},
  \bibinfo{author}{\bibfnamefont{W.~N.} \bibnamefont{Mei}},
  \bibinfo{author}{\bibfnamefont{P.~A.} \bibnamefont{Dowben}},
  \bibinfo{author}{\bibfnamefont{S.~S.} \bibnamefont{Jaswal}},
  \bibnamefont{and} \bibinfo{author}{\bibfnamefont{E.~Y.}
  \bibnamefont{Tsymbal}}, \bibinfo{journal}{J.\ Phys.: Condens.\ Matter}
  \textbf{\bibinfo{volume}{19}}, \bibinfo{pages}{315220}
  (\bibinfo{year}{2007}).

\bibitem[{\citenamefont{Chantis et~al.}(2007)\citenamefont{Chantis, van
  Schilfgaarde, and Kotani}}]{Chantis_Kotani}
\bibinfo{author}{\bibfnamefont{A.~N.} \bibnamefont{Chantis}},
  \bibinfo{author}{\bibfnamefont{M.}~\bibnamefont{van Schilfgaarde}},
  \bibnamefont{and} \bibinfo{author}{\bibfnamefont{T.}~\bibnamefont{Kotani}},
  \bibinfo{journal}{Phys. Rev. B} \textbf{\bibinfo{volume}{76}},
  \bibinfo{pages}{165126} (\bibinfo{year}{2007}).

\bibitem[{\citenamefont{Mitra and
  Lambrecht}(2008{\natexlab{a}})}]{Mitra_Lambrecht2}
\bibinfo{author}{\bibfnamefont{C.}~\bibnamefont{Mitra}} \bibnamefont{and}
  \bibinfo{author}{\bibfnamefont{W.~R.~L.} \bibnamefont{Lambrecht}},
  \bibinfo{journal}{Phys.\ Rev.\ B} \textbf{\bibinfo{volume}{78}},
  \bibinfo{pages}{195203} (\bibinfo{year}{2008}{\natexlab{a}}).

\bibitem[{\citenamefont{Trodahl et~al.}(2007)\citenamefont{Trodahl, Preston,
  Zhong, Ruck, Strickland, Mitra, and Lambrecht}}]{Trodahl_Lambrecht}
\bibinfo{author}{\bibfnamefont{H.~J.} \bibnamefont{Trodahl}},
  \bibinfo{author}{\bibfnamefont{A.~R.~H.} \bibnamefont{Preston}},
  \bibinfo{author}{\bibfnamefont{J.}~\bibnamefont{Zhong}},
  \bibinfo{author}{\bibfnamefont{B.~J.} \bibnamefont{Ruck}},
  \bibinfo{author}{\bibfnamefont{N.}~\bibnamefont{Strickland}},
  \bibinfo{author}{\bibfnamefont{C.}~\bibnamefont{Mitra}}, \bibnamefont{and}
  \bibinfo{author}{\bibfnamefont{W.~R.~L.} \bibnamefont{Lambrecht}},
  \bibinfo{journal}{Phys. Rev. B} \textbf{\bibinfo{volume}{76}},
  \bibinfo{pages}{085211} (\bibinfo{year}{2007}).

\bibitem[{\citenamefont{Yoshitomi
  et~al.}(2011{\natexlab{a}})\citenamefont{Yoshitomi, Kitayama, Kita, Wada,
  Fujisawa, Ohta, and Sakurai}}]{Yoshitomi_Sakurai}
\bibinfo{author}{\bibfnamefont{H.}~\bibnamefont{Yoshitomi}},
  \bibinfo{author}{\bibfnamefont{S.}~\bibnamefont{Kitayama}},
  \bibinfo{author}{\bibfnamefont{T.}~\bibnamefont{Kita}},
  \bibinfo{author}{\bibfnamefont{O.}~\bibnamefont{Wada}},
  \bibinfo{author}{\bibfnamefont{M.}~\bibnamefont{Fujisawa}},
  \bibinfo{author}{\bibfnamefont{H.}~\bibnamefont{Ohta}}, \bibnamefont{and}
  \bibinfo{author}{\bibfnamefont{T.}~\bibnamefont{Sakurai}},
  \bibinfo{journal}{Phys. Rev. B} \textbf{\bibinfo{volume}{83}},
  \bibinfo{pages}{155202} (\bibinfo{year}{2011}{\natexlab{a}}).

\bibitem[{\citenamefont{Yoshitomi
  et~al.}(2011{\natexlab{b}})\citenamefont{Yoshitomi, Kitayama, Kita, Wada,
  Fujisawa, Ohta, and Sakurai}}]{Yoshitomi_Sakurai2}
\bibinfo{author}{\bibfnamefont{H.}~\bibnamefont{Yoshitomi}},
  \bibinfo{author}{\bibfnamefont{S.}~\bibnamefont{Kitayama}},
  \bibinfo{author}{\bibfnamefont{T.}~\bibnamefont{Kita}},
  \bibinfo{author}{\bibfnamefont{O.}~\bibnamefont{Wada}},
  \bibinfo{author}{\bibfnamefont{M.}~\bibnamefont{Fujisawa}},
  \bibinfo{author}{\bibfnamefont{H.}~\bibnamefont{Ohta}}, \bibnamefont{and}
  \bibinfo{author}{\bibfnamefont{T.}~\bibnamefont{Sakurai}},
  \bibinfo{journal}{Phys. Status Solidi C} \textbf{\bibinfo{volume}{8}},
  \bibinfo{pages}{488} (\bibinfo{year}{2011}{\natexlab{b}}).

\bibitem[{\citenamefont{Preston et~al.}(2010)\citenamefont{Preston, Ruck,
  Lambrecht, Piper, Downes, and Trodahl}}]{Preston_Trodahl}
\bibinfo{author}{\bibfnamefont{A.~R.~H.} \bibnamefont{Preston}},
  \bibinfo{author}{\bibfnamefont{B.~J.} \bibnamefont{Ruck}},
  \bibinfo{author}{\bibfnamefont{W.~R.~L.} \bibnamefont{Lambrecht}},
  \bibinfo{author}{\bibfnamefont{L.~F.~J.} \bibnamefont{Piper}},
  \bibinfo{author}{\bibfnamefont{J.~E.} \bibnamefont{Downes}},
  \bibnamefont{and} \bibinfo{author}{\bibfnamefont{K.~E. S. H.~J.}
  \bibnamefont{Trodahl}}, \bibinfo{journal}{Appl. Phys. Lett.}
  \textbf{\bibinfo{volume}{96}}, \bibinfo{pages}{032101}
  (\bibinfo{year}{2010}).

\bibitem[{\citenamefont{Senapati
  et~al.}(2011{\natexlab{b}})\citenamefont{Senapati, Blamire, and
  Barber}}]{Senapati_Barber2}
\bibinfo{author}{\bibfnamefont{K.}~\bibnamefont{Senapati}},
  \bibinfo{author}{\bibfnamefont{M.~G.} \bibnamefont{Blamire}},
  \bibnamefont{and} \bibinfo{author}{\bibfnamefont{Z.~H.}
  \bibnamefont{Barber}}, \bibinfo{journal}{Nature Materials}
  \textbf{\bibinfo{volume}{10}}, \bibinfo{pages}{849}
  (\bibinfo{year}{2011}{\natexlab{b}}).

\bibitem[{\citenamefont{Kasuya}(1970)}]{Kasuya}
\bibinfo{author}{\bibfnamefont{T.}~\bibnamefont{Kasuya}},
  \bibinfo{journal}{I.B.M.~J.\ Res.\ Develop.} \textbf{\bibinfo{volume}{14}},
  \bibinfo{pages}{214} (\bibinfo{year}{1970}).

\bibitem[{\citenamefont{Lee and Liu}(1984)}]{Lee_Liu}
\bibinfo{author}{\bibfnamefont{V.-C.} \bibnamefont{Lee}} \bibnamefont{and}
  \bibinfo{author}{\bibfnamefont{L.}~\bibnamefont{Liu}},
  \bibinfo{journal}{Phys.\ Rev.\ B} \textbf{\bibinfo{volume}{30}},
  \bibinfo{pages}{2026} (\bibinfo{year}{1984}).

\bibitem[{\citenamefont{Kasuya and Li}(1997)}]{Kasuya_Li}
\bibinfo{author}{\bibfnamefont{T.}~\bibnamefont{Kasuya}} \bibnamefont{and}
  \bibinfo{author}{\bibfnamefont{D.~X.} \bibnamefont{Li}},
  \bibinfo{journal}{J.\ Magn.\ Magn.\ Mater.} \textbf{\bibinfo{volume}{167}},
  \bibinfo{pages}{L1} (\bibinfo{year}{1997}).

\bibitem[{\citenamefont{Mitra and
  Lambrecht}(2008{\natexlab{b}})}]{Mitra_Lambrecht}
\bibinfo{author}{\bibfnamefont{C.}~\bibnamefont{Mitra}} \bibnamefont{and}
  \bibinfo{author}{\bibfnamefont{W.~R.~L.} \bibnamefont{Lambrecht}},
  \bibinfo{journal}{Phys.\ Rev.\ B} \textbf{\bibinfo{volume}{78}},
  \bibinfo{pages}{134421} (\bibinfo{year}{2008}{\natexlab{b}}).

\bibitem[{\citenamefont{Wachter and Kaldis}(1980)}]{Wachter_Kaldis}
\bibinfo{author}{\bibfnamefont{P.}~\bibnamefont{Wachter}} \bibnamefont{and}
  \bibinfo{author}{\bibfnamefont{E.}~\bibnamefont{Kaldis}},
  \bibinfo{journal}{Solid State Commun.} \textbf{\bibinfo{volume}{34}},
  \bibinfo{pages}{241} (\bibinfo{year}{1980}).

\bibitem[{\citenamefont{Wachter}(2012)}]{Wachter}
\bibinfo{author}{\bibfnamefont{P.}~\bibnamefont{Wachter}},
  \bibinfo{journal}{Results in Phys.} \textbf{\bibinfo{volume}{2}},
  \bibinfo{pages}{90} (\bibinfo{year}{2012}).

\bibitem[{\citenamefont{Duan et~al.}(2006)\citenamefont{Duan, Sabiryanov, Mei,
  Dowben, Jaswal, and Tsymbal}}]{Duan_Tsymbal2}
\bibinfo{author}{\bibfnamefont{C.-G.} \bibnamefont{Duan}},
  \bibinfo{author}{\bibfnamefont{R.~F.} \bibnamefont{Sabiryanov}},
  \bibinfo{author}{\bibfnamefont{W.~N.} \bibnamefont{Mei}},
  \bibinfo{author}{\bibfnamefont{P.~A.} \bibnamefont{Dowben}},
  \bibinfo{author}{\bibfnamefont{S.~S.} \bibnamefont{Jaswal}},
  \bibnamefont{and} \bibinfo{author}{\bibfnamefont{E.~Y.}
  \bibnamefont{Tsymbal}}, \bibinfo{journal}{Appl.\ Phys.\ Lett.}
  \textbf{\bibinfo{volume}{88}}, \bibinfo{pages}{182505}
  (\bibinfo{year}{2006}).

\bibitem[{\citenamefont{Sharma and Nolting}(2010)}]{Sharma_Nolting}
\bibinfo{author}{\bibfnamefont{A.}~\bibnamefont{Sharma}} \bibnamefont{and}
  \bibinfo{author}{\bibfnamefont{W.}~\bibnamefont{Nolting}},
  \bibinfo{journal}{Phys.\ Rev.\ B} \textbf{\bibinfo{volume}{81}},
  \bibinfo{pages}{125303} (\bibinfo{year}{2010}).

\bibitem[{\citenamefont{Plank et~al.}(2011)\citenamefont{Plank, Natali,
  Galipaud, Richter, Simpson, Trodahl, and Ruck}}]{Plank_Ruck}
\bibinfo{author}{\bibfnamefont{N.~O.~V.} \bibnamefont{Plank}},
  \bibinfo{author}{\bibfnamefont{F.}~\bibnamefont{Natali}},
  \bibinfo{author}{\bibfnamefont{J.}~\bibnamefont{Galipaud}},
  \bibinfo{author}{\bibfnamefont{J.}~\bibnamefont{Richter}},
  \bibinfo{author}{\bibfnamefont{M.}~\bibnamefont{Simpson}},
  \bibinfo{author}{\bibfnamefont{H.~J.} \bibnamefont{Trodahl}},
  \bibnamefont{and} \bibinfo{author}{\bibfnamefont{B.~J.} \bibnamefont{Ruck}},
  \bibinfo{journal}{Appl. Phys. Lett.} \textbf{\bibinfo{volume}{98}},
  \bibinfo{pages}{112503} (\bibinfo{year}{2011}).

\bibitem[{\citenamefont{Senapati et~al.}(2010)\citenamefont{Senapati, Fix,
  Vickers, Blamire, and Barber}}]{Senapati_Barber3}
\bibinfo{author}{\bibfnamefont{K.}~\bibnamefont{Senapati}},
  \bibinfo{author}{\bibfnamefont{T.}~\bibnamefont{Fix}},
  \bibinfo{author}{\bibfnamefont{M.~E.} \bibnamefont{Vickers}},
  \bibinfo{author}{\bibfnamefont{M.~G.} \bibnamefont{Blamire}},
  \bibnamefont{and} \bibinfo{author}{\bibfnamefont{Z.~H.}
  \bibnamefont{Barber}}, \bibinfo{journal}{J. Phys.: Condens. Matter}
  \textbf{\bibinfo{volume}{22}}, \bibinfo{pages}{302003}
  (\bibinfo{year}{2010}).

\bibitem[{Mvs()}]{MvsTSamples}
\bibinfo{note}{Sample $L$ was grown using the methodology described in
  Ref.~[\onlinecite{Granville_Trodahl}] but under a higher N$_2$ partial
  pressure of $4\times10^{-4}$~mbar. Sample $M$ is the present work, and the
  growth conditions of sample $H$ can be found in
  Ref.~[\onlinecite{Natali_Hirsch}].}

\bibitem[{\citenamefont{Natali et~al.}(2010)\citenamefont{Natali, Plank,
  Galipaud, Ruck, Trodahl, Semond, Sorieul, and Hirsch}}]{Natali_Hirsch}
\bibinfo{author}{\bibfnamefont{F.}~\bibnamefont{Natali}},
  \bibinfo{author}{\bibfnamefont{N.~O.~V.} \bibnamefont{Plank}},
  \bibinfo{author}{\bibfnamefont{J.}~\bibnamefont{Galipaud}},
  \bibinfo{author}{\bibfnamefont{B.~J.} \bibnamefont{Ruck}},
  \bibinfo{author}{\bibfnamefont{H.~J.} \bibnamefont{Trodahl}},
  \bibinfo{author}{\bibfnamefont{F.}~\bibnamefont{Semond}},
  \bibinfo{author}{\bibfnamefont{S.}~\bibnamefont{Sorieul}}, \bibnamefont{and}
  \bibinfo{author}{\bibfnamefont{L.}~\bibnamefont{Hirsch}},
  \bibinfo{journal}{J. Cryst. Growth} \textbf{\bibinfo{volume}{312}},
  \bibinfo{pages}{3583} (\bibinfo{year}{2010}).

\bibitem[{\citenamefont{Blundell}(2001)}]{Blundell}
\bibinfo{author}{\bibfnamefont{S.}~\bibnamefont{Blundell}},
  \emph{\bibinfo{title}{Magnetism in Condensed Matter}}
  (\bibinfo{publisher}{Oxford University Press}, \bibinfo{year}{2001}).

\bibitem[{\citenamefont{Fisher and Langer}(1968)}]{Fisher_Langer}
\bibinfo{author}{\bibfnamefont{M.~E.} \bibnamefont{Fisher}} \bibnamefont{and}
  \bibinfo{author}{\bibfnamefont{J.~S.} \bibnamefont{Langer}},
  \bibinfo{journal}{Phys.\ Rev.\ Lett.} \textbf{\bibinfo{volume}{20}},
  \bibinfo{pages}{665} (\bibinfo{year}{1968}).

\bibitem[{\citenamefont{Punya et~al.}(2010)\citenamefont{Punya,
  Cheiwchanchamnangij, Thiess, and Lambrecht}}]{Punya_Lambrecht}
\bibinfo{author}{\bibfnamefont{A.}~\bibnamefont{Punya}},
  \bibinfo{author}{\bibfnamefont{T.}~\bibnamefont{Cheiwchanchamnangij}},
  \bibinfo{author}{\bibfnamefont{A.}~\bibnamefont{Thiess}}, \bibnamefont{and}
  \bibinfo{author}{\bibfnamefont{W.~R.~L.} \bibnamefont{Lambrecht}},
  \bibinfo{journal}{Mater. Res. Soc. Symp. Proc.}
  \textbf{\bibinfo{volume}{1290}}, \bibinfo{pages}{10.1557/opl.2011.383}
  (\bibinfo{year}{2010}).

\bibitem[{\citenamefont{Coey et~al.}(2005)\citenamefont{Coey, Venkatesan, and
  Fitzgerald}}]{Coey_Venkatesan}
\bibinfo{author}{\bibfnamefont{J.~M.~D.} \bibnamefont{Coey}},
  \bibinfo{author}{\bibfnamefont{M.}~\bibnamefont{Venkatesan}},
  \bibnamefont{and} \bibinfo{author}{\bibfnamefont{C.~B.}
  \bibnamefont{Fitzgerald}}, \bibinfo{journal}{Nature Materials}
  \textbf{\bibinfo{volume}{4}}, \bibinfo{pages}{173} (\bibinfo{year}{2005}).

\bibitem[{\citenamefont{Azeem}()}]{AzeemPComm}
\bibinfo{author}{\bibfnamefont{M.}~\bibnamefont{Azeem}}, \bibinfo{note}{private
  communication}.

\bibitem[{\citenamefont{de~Gennes and Friedel}(1958)}]{deGennes_Friedel}
\bibinfo{author}{\bibfnamefont{P.}~\bibnamefont{de~Gennes}} \bibnamefont{and}
  \bibinfo{author}{\bibfnamefont{J.}~\bibnamefont{Friedel}},
  \bibinfo{journal}{J.\ Phys.\ Chem.\ Solids} \textbf{\bibinfo{volume}{4}},
  \bibinfo{pages}{71} (\bibinfo{year}{1958}).

\end{thebibliography}

\end{document}